\documentclass[aps,prd,reprint]{revtex4-2}
\usepackage{comment}

\usepackage{amsmath,amssymb,hyperref,mathrsfs}
\usepackage{braket,bm,bbm}
\PassOptionsToPackage{normalem}{ulem}

\usepackage[makeroom]{cancel}
\usepackage{graphicx}
\usepackage{amsthm}
\usepackage[usenames,dvipsnames]{xcolor}
\hypersetup{colorlinks=true, citecolor=Plum, linkcolor=Plum,
urlcolor=Plum}


\usepackage{subfigure}
\begin{document}
    
\title{A Singularity Theorem for Quasi-Topological Gravities}
\author{Francesco Di Filippo}
\affiliation{Institut f\"ur Theoretische Physik, Max-von-Laue-Str.1, 60438 Frankfurt, Germany\\ $\,$}

\begin{abstract}
 \noindent Quasi-topological theories of gravity have recently attracted substantial attention as several theories in this class have a non-singular black hole as a vacuum and spherically symmetric solution. This motivates the expectation that these theories can provide a resolution to the singularity problem and a description of black holes cores. The known solutions, however, possess a Cauchy horizon which denotes the end of predictability of the theory, as the initial data do not determine a unique continuation past this type of horizon. In this paper, we show that Penrose singularity theorem can be extended to all the known classes of quasi-topological theories that have non-singular black holes as solutions. We prove that spherical collapse satisfying the null energy condition necessarily leads to future null geodesic incompleteness once a trapped sphere forms. This implies that the lack of predictability of the theories is not an artefact of vacuum solutions and these theories are as problematic as general relativity for the dynamical studies of black hole interiors.
\end{abstract}

\maketitle
\section{Introduction}

Black holes singularities represent one of the most fascinating open problems in the studies of fundamental physics. The definition of what a singularity is can sometimes be elusive~\cite{Geroch:1968ut,Carballo-Rubio:2025fnc}. Singularities, are often associated with the blowing up of curvature invariants. However, divergences in the curvature invariants are not in one-to-one relation with the end of predictability of the theory. A more systematic approach consists in defining singularities as the failure of the theory to be predictive. With this definition of singularities, there are a series of singularity theorems proving that in general relativity the gravitational collapse of matter satisfying suitable energy conditions leads a singular spacetime, i.e. a geodesically incomplete spacetime~\cite{Penrose:1964wq,hawking1973,Senovilla:1998oua}. 

This has led to an intensive program to investigate the nature of black hole interiors to make sense of singularities. Remarkably, several aspects of the problem can be studied in a theory independent way~\cite{Carballo-Rubio:2018pmi,Carballo-Rubio:2019fnb}. However, it is necessary to specify the framework to fully understand the physics at play. Recently, quasi-topological theories of gravity have gained a lot of popularity. Such theories are higher dimensional theories that have the peculiarities of having very simple equation of motion once restricted to spherically symmetric spacetime. It was recently shown the remarkable result that several quasi-topological theories have regular black holes as vacuum solutions~\cite{Oliva:2010eb,Myers:2010ru,Bueno:2024dgm,Bueno:2025qjk}. This result was then extended to 4-dimensional spacetimes with the introduction of non-polynomial gravities~\cite{Bueno:2025zaj}. Therefore, these theories can resolve the problems related to the divergences of curvature invariants. However, all the solutions obtained contain a Cauchy horizon which implies the end of predictability of the theory. Therefore, the maximal Cauchy development of the know solutions are geodesically incomplete. In this paper we show that the problem of geodesically incompleteness is not only relative to vacuum solutions but it is a general issue with these theories. In fact, we show that Penrose's incompleteness theorem can be extend to be applicable to all the quasi-topological theories which are known to provide regular black holes as vacuum solutions.

Therefore, the (spherically symmetric) gravitational collapse in quasi-topological theories of gravity of matter satisfying the null energy condition is as problematic as the gravitational collapse of the same matter distribution in general relativity. 

\section{Penrose Theorem in General Relativity}

Let us start by reviewing the usual formulation of Penrose theorem. 

\paragraph*{Penrose incompleteness theorem}
Let $(\mathcal{M},g)$ be a smooth and time orientable Lorenzian manifold such that 
\begin{itemize}
\item[\textit{i.}] The spacetime is globally hyperbolic with non-compact Cauchy hypersurfaces;
\item[\textit{ii.}] There is a trapped region $\mathcal{T}$;
\item[\textit{iii.}] The null convergence $R_{ab}k^a k^b\geq0$ conditions are satisfied in the future of $\mathcal{T}$.
\end{itemize}
Then
\textit{The spacetime is future null geodesically incomplete}

Besides the standard formulation of the theorem, we will follow the logic of the \textit{one-sided} Penrose incompleteness theorem~\cite{Andersson:2008up,Galloway:2024jwi}. 

\paragraph*{One-sided Penrose incompleteness theorem} Let $(\mathcal{M},g)$ be a smooth and time orientable Lorenzian manifold such that 
\begin{itemize}
\item[\textit{i.}] The spacetime is globally hyperbolic with non-compact Cauchy hypersurfaces;
\item[\textit{ii.}]  There is a smooth, compact, connected two-sided surface
$T$ in a Cauchy hypersurface $\Sigma$, separating $\Sigma$ into two components, such that $\theta_{\rm out}<0$ for the future null normal pointing towards a side of $\Sigma$  with
non-compact closure.
\item[\textit{iii.}] The null convergence conditions $R_{ab}k^a k^b\geq0$ are satisfied at each point in the future of $\mathcal{T}$ along the outgoing null geodesic. 
\end{itemize}
Then
\textit{The spacetime is future null geodesically incomplete}
We can see that this version of the theorem only imposes that $\mathcal{T}$ is outer trapped and that the null convergence conditions are satisfied along the outgoing null direction.

As it is clear from the assumoptions of the theorems, there is no reference to specific field equation of the theories. 

The key ingredient in both proofs is the focusing of null geodesics orthogonal to the trapped surface. For an affinely
parametrized null congruence with tangent $k^a$, its expansion is the fractional rate of change of the transverse area
element,
\begin{equation}
    \theta=\frac{1}{\delta A}
           \frac{\mathrm{d}(\delta A)}{\mathrm{d}\lambda}.
\end{equation}
For radial congruences in spherical symmetry, this reduces to
\begin{equation}\label{eq:def_exp}
    \theta=\frac{D-2}{r}\frac{\mathrm{d}r}{\mathrm{d}\lambda}.
\end{equation}
The congruences orthogonal to the surface have vanishing
twist. Their expansion obeys the Raychaudhuri equation,
\begin{equation}
    \frac{\mathrm{d}\theta}{\mathrm{d}\lambda}
    =-\frac{\theta^2}{D-2}
     -\sigma_{ab}\sigma^{ab}
     -R_{ab}k^ak^b,
\end{equation}
where $\sigma_{ab}$ is the shear. Null convergence therefore
implies
\begin{equation}
    \frac{\mathrm{d}\theta}{\mathrm{d}\lambda}
    \leq-\frac{\theta^2}{D-2},
\end{equation}
so an initially negative expansion diverges to $-\infty$
within finite affine parameter, provided the generator extends that far. The resulting focal points prevent the generators from remaining on the achronal future boundary.
Global hyperbolicity and the relevant non-compactness assumption then yield a contradiction if future null completeness is assumed. In the one-sided version, focusing is required only for the outgoing generators.

The theories considered here employ the torsion-free Levi--Civita connection, so the geodesic deviation and Raychaudhuri equations retain their standard geometric form. The modification enters only through the field equations relating curvature to matter. The purely geometric nature of Penrose's singularity theorems implies that they trivially apply to quasi-topological theories. 

However, one of the assumption of the theorem is the null convergence condition. In general relativity this is a reasonable assumption as it is equivalent to the null energy condition via the Einstein equations
\begin{equation}
        T_{ab}k^a k^b\geq0\,,\quad\iff\quad R_{ab}k^a k^b\geq0\,.
\end{equation}
Therefore, assuming the null convergence conditions would not necessarily be justified if the Einstein equations do not hold. We will assume the null energy conditions and additional conditions whose validity we will discuss in the next sections.

\section{Penrose theorem for quasi-topological gravities}

We consider two classes of higher-curvature metric theories: polynomial Birkhoff quasi-topological gravities in $D\geq5$ and their four-dimensional non-polynomial counterparts constructed whose spherical field equations take the form \eqref{eq:qtg_equations}. More general higher-curvature theories admitting regular black holes \cite{Carballo-Rubio:2025ntd} lie outside the scope of our analysis. Although their covariant actions differ, their field equations on general, time-dependent, spherically symmetric metrics have the same form. We therefore treat both classes simultaneously.

Consider a general spherically symmetric metric,
\begin{equation}
    ds^2
    =\gamma_{\mu\nu}(x)\,dx^\mu dx^\nu
     +r(x)^2d\Omega_{D-2}^2 ,
\end{equation}
where the Greek indices denote denote the 2-dimensional quantities orthogonal to the $D-2$ spheres.
Let us define
\begin{equation}\label{eq:definitions_qtg}
    \begin{array}{c}
   \displaystyle    X\equiv \gamma^{\mu\nu}\nabla_\mu r \nabla_\nu r,
    \qquad
    \psi\equiv\frac{1-X}{r^2},
    \\
    \\
   \displaystyle h(\psi)
    \equiv
    \psi+\sum_{n=2}^{\infty}
    c_n\psi^n .
\end{array}
\end{equation}
with $c_n$ being coefficients that depends on the coupling constants of the theory
The field equations of quasi-topological gravities take the form~\cite{Bueno:2024zsx,Bueno:2025zaj}
\begin{equation}\label{eq:qtg_equations}
    -\frac{D-2}{r}h'(\psi)\nabla_\mu \nabla_\nu r
    +\mathcal{B}\gamma_{\mu\nu}
    =8\pi G_N T_{\mu\nu},
\end{equation}
where $\mathcal{B}$ is defined in~\cite{Bueno:2024zsx}.

With these elements it is possible to prove the following\\\\
\noindent\textit{Theorem 1.} Consider a spherically symmetric Lorenzian maninfold $(\mathcal{M},g)$ which is a smooth, non-compact, and time-oriented solution of the quasi-topological theories such that 
\begin{enumerate}
    \item[\textit{i.}] The spacetime is globally hyperbolic with non-compact Cauchy hypersurfaces;
    \item[\textit{ii.}] There is a trapped spherical surface $\mathcal{T}$;
    \item[\textit{iii.}] The null energy condition $        T_{\mu\nu}k^\mu k^\nu>0\,,$ holds  $\forall k^\mu $ null;
    \item[\textit{iv.}] $h'(\psi(p))>0$ $\forall p\in\mathcal{M}\cap J^+(\mathcal{T})$.    
\end{enumerate}
Then $\mathcal{M}$ is null geodesically incomplete.\\
\begin{proof}
Conditions \textit{i.}~and \textit{ii.}~ are the same as in Penrsose theorem. Let us prove that the conditions \textit{iii.}~and \textit{iv.}~ imply the null convergence conditions. 

Let $k^a=(k^\mu,0)$ be a radial null vector, so that
$\gamma_{\mu\nu}k^\mu k^\nu=0$. Contracting the field
equations with $k^\mu k^\nu$, all terms proportional
to $\gamma_{\mu\nu}$ vanish:
\begin{align}
    8\pi G_N T_{\mu\nu}k^\mu k^\nu
    &= -\frac{D-2}{ r}\,h'(\psi)\,
       k^\mu k^\nu\nabla_\mu\nabla_\nu r.
\end{align}

Independently of the field equations, the warped-product
Ricci identity gives
\begin{equation}
    R^{(D)}_{\mu\nu}
    = R^{(\gamma)}_{\mu\nu}
      - \frac{D-2}{ r}
        \nabla_\mu\nabla_\nu r.
\end{equation}
Here the superscripts distinguish the full spacetime
Ricci tensor from that of $\gamma_{\mu\nu}$. In two
dimensions,
\begin{equation}
    R^{(\gamma)}_{\mu\nu}
    = \frac12 R^{(\gamma)}\gamma_{\mu\nu},
\end{equation}
and therefore its contraction with $k^\mu k^\nu$
vanishes. Consequently,
\begin{equation}
    R^{(D)}_{\mu\nu}k^\mu k^\nu
    = -\frac{D-2}{ r}\,
      k^\mu k^\nu\nabla_\mu\nabla_\nu r.
\end{equation}
Combining the two identities and using the fact that $k^a$ does not have angular components,
\begin{equation}
        h'(\psi)\,R^{(D)}_{ab}k^a k^b
        = 8\pi G_N T_{ab}k^a k^b\,.
\end{equation}

Thus, because  $h'(\psi)$ is  positive by assumption \textit{iv.}, the matter null energy condition implies the radial null convergence condition:
\begin{equation}
    R^{(D)}_{ab}k^a k^b
    = \frac{8\pi G_N}{h'(\psi)}T_{ab}k^a k^b
    \geq 0.
\end{equation}
At this point, all the steps of Penrose's original theorem can be followed to prove null geodesic incompleteness.
\end{proof}

\section{Validity of the assumptions}

We have seen that Penrose's theorem can be extended to quasi-topological theories of gravity provided $h'(\psi)\geq0$. However, we have to quantify how restrictive this condition is.
From the definition of $h(\psi)$ in Eq.~\eqref{eq:definitions_qtg} it immediately follows
\begin{equation}
    h'(\psi)=1+\sum_{n=2}^\infty nc_n\psi^{n-1}\,,
\end{equation}
where, let us remind, the coefficient $c_n$ depends on the specific theory under consideration, not on the specific solution.

In Refs.~\cite{Bueno:2024dgm,Bueno:2025zaj} the authors impose the conditions $c_n\geq 0$ as part of a sufficient condition to ensure convergence of the series that defines $h(\psi)$. Under these condition, we have
\begin{equation}\label{eq:positivity}
    h'(\psi)\geq1\qquad\text{for}\qquad \psi\geq 0\,.
\end{equation}
To the author's knowledge, the only theory known to have non-singular black holes solutions for which some of the $c_n$ coefficient are negative is given in \cite{DiFilippo:2024mwm}. However, the result of this work also applies to that theory as the condition in Eq.~\eqref{eq:positivity} is satisfied in that theory.

The condition \eqref{eq:positivity} is weaker than the one we need to simply apply the result of the previous section. However, we will see that this is enough to prove a refined version of the theorem. 
\section{Refined theorem}
In this section, we are going to prove a refined version of the theorem in which we only need to assume the non-negativity of $h'(\psi)$ for positive $\psi$. As seen in the previous section, this would include all the known quasi-topological theories that are known to have regular black holes as vacuum solution.

We can formulate the following\\\\
\textit{Theorem 2.}~Consider a spherically symmetric Lorenzian maninfold $(\mathcal{M},g)$ which is a smooth, non-compact, and time-oriented classical solution of the quasi-topological theories such that 
\begin{enumerate}
    \item[\textit{i.}] The spacetime is globally hyperbolic with non-compact Cauchy hypersurfaces;
    \item[\textit{ii.}] The ingoing null expansion $\theta_{in}$ is negative at any point of an  initial Cauchy slice $\Sigma_0$; 
    \item[\textit{iii.}] There is a closed spherical surface $\mathcal{T}$ which is outer trapped $\theta_{out}<0$;
    \item[\textit{iv.}] The null energy condition $ T_{\mu\nu}k^\mu k^\nu\geq0\,,$ holds  $\forall k^\mu $ null;
    \item[\textit{v.}] $h'(\psi(p))>0$ for $ \psi>0$.    
\end{enumerate}
Then $\mathcal{M}$ is null geodesically incomplete.\\

Before proving the theorem, let us discuss the validity of the assumptions. Condition \textit{i.} is the usual condition required to have a well defined initial value problem.
condition \textit{ii.} is not assumed in the standard Penrose theorem, however, note that it is only assumed to hold at arbitrary early times when the spacetime in a realistic gravitational collapse is arbitrarily close to Minkowski, where this condition holds. Condition \textit{iii.} is implied by the presence of a trapped region. Condition \textit{iv.} is the same energy condition imposed on the matter distribution in general relativity. Finally, condition \textit{v.} is satisfied for all the quasi-topological theories of gravity known to have regular black holes as vacuum solutions. 

 Of course, we need to precisely define what we mean by ingoing and outgoing null direction. 

Denote $\Sigma$ the Cauchy slice which includes $\mathcal{T}$. As $\mathcal{T}$ is a closed (D-2)-sphere, it divides the Cauchy hypersurface into two sides. We can write $\Sigma\setminus\mathcal{T}=\Sigma^-\cup \Sigma^+$ with $\bar{\Sigma}^+$ is non-compact (because $\Sigma$ is non compact, at least one between $\bar{\Sigma}^-$ and $\bar{\Sigma}^+$ must be non-compact). The outgoing null direction is defined as the one whose projection on $\Sigma$ points towards $\Sigma^+$. In practice, the outgoing null direction is the one that does not point towards the center of the spacetime\footnote{For spacetimes like wormholes in which both sides are non-compact, the theorem can be applied by defining either null direction as the outgoing one.}. 

The logic of the proof goes as follow. First of all, we will show that $\psi>0$ inside a trapped region. In turn, this implies that $h'(\psi(p))>0$ there.  This is not enough to prove the refined version of the theorem. While the initial conditions are given inside the trapped region, null geodesics can in principle escape it (and the ingoing ones generally do leave it).  
The next step of the proof will be to show that under the assumption of the theorem the outgoing null geodesics cannot leave a trapped region. This implies that the null convergence conditions hold everywhere in the future of the trapped region along the outgoing null geodesic. Therefore, we can trivially apply the one-sided version of the Penrose theorem to prove the wanted result.

To improve clarity, we will prove the crucial steps as separate lemmas.\\

\noindent\textit{Lemma 1. Under the assumptions of the theorem, $h'(\psi)> 0$ inside a trapped region. }
\begin{proof}
We are going to show that inside the trapped region $\psi>0$ and hence $h'(\psi)> 0$. This is very easily done by looking at the definition of $\psi$ in Eq.~\eqref{eq:definitions_qtg}. 
Introducing the two independent radial vector fields $\ell$ and $k$ such has $\ell\cdot k=-1$, we have
\begin{equation}
    \gamma^{\mu\nu}=- \ell^\mu k^\nu- \ell^\nu k^\mu\,.
\end{equation}
Therefore,
\begin{equation}\label{eq:Nablar}
    \gamma^{\mu\nu}\nabla_\mu r\nabla_\nu r=-2\ell^\nu k^\mu\nabla_\mu r\nabla_\nu r=-2\frac{dr}{d\lambda_{in}}\frac{dr}{d\lambda_{out}}\,,
\end{equation}
As seen in Eq.~\eqref{eq:def_exp}, the variation of the area radius along the affine parameter is proportional to the expansion parameter of such geodesic. Thus, inside the trapped region, the variation of the area radius along the affine parameter of both independent null geodesic is negative. Therefore
\begin{equation}
    \gamma^{\mu\nu}\nabla_\mu r\nabla_\nu r<0\,\implies\, \psi=\frac{1-\gamma^{\mu\nu}\nabla_\mu r\nabla_\nu r}{r^2}>\frac{1}{r^2}>0\,.
\end{equation}
Therefore, inside the trapped region $\psi$ is positive and, by assumption \textit{v.}, so is $h'(\psi)$.
\end{proof}

\noindent\textit{Lemma 2. Under the assumptions of the theorem, an initially negative expansion parameter remains
strictly negative along the generator of the corresponding geodesic unless it reaches the center of symmetry $r=0$.}
\begin{proof}
By assumption
\begin{equation}
   \left. \frac{dr}{d\lambda}\right|_{\lambda=\lambda_0}<0\,.
\end{equation}
Suppose the expansion first vanish at $\lambda_1$ the first value of the affine parameter at which the expansion vanishes. 
\begin{equation}
     \left. \frac{dr}{d\lambda}\right|_{\lambda=\lambda_1}=0\qquad\text{and}\qquad\frac{dr}{d\lambda} <0\quad \text{for }\lambda_0<\lambda<\lambda_1\,.
\end{equation}
At $\lambda_1$, from Eq.~\eqref{eq:Nablar} we have
\begin{equation}
    g^{\mu\nu}\nabla_\mu r\nabla_\nu r=0\quad\implies\quad \psi=\frac{1}{r^2}>0\,.
\end{equation}
By continuity, there must be an interval centered in $\lambda_1$ in which $\psi$ is positive. Therefore, by assumption, the null convergence conditions are satisfied there.  Considering a value $\bar{\lambda}<\lambda_1$ in this interval, we  know that the null convergence conditions are satisfied. Therefore, the Raychaudhuri equation implies that the derivative of the expansion is negative. We also know that the expansion must be negative as the first zero of the expansion parameter is $\lambda_1$ by assumption. Therefore, the expansion has to decrease and cannot reach zero at $\lambda_1$. Thus we reached a contradiction proving the desired result that expansion parameter cannot change sign. 
\end{proof} 

\noindent\textit{Lemma 3.
Under the assumptions of the theorem, the ingoing null
expansion satisfies $\theta_{\rm in}<0$ everywhere in the future of $\Sigma_0$.}
\begin{proof}
Let us consider a generic point $p$ in the future of $\Sigma_0$and show that $\theta_{in}<0$. 
Trace the ingoing radial null geodesic through $p$ toward the past. By global hyperbolicity, tracing back any null geodesic we must reach $\Sigma_0$, unless it first encounters $r=0$, where the radial ingoing null geodesics become radial outgoing and vice-versa.  By  definition of the ingoing direction, tracing the geodesic toward the past moves it away from the center.
The ingoing expansion is negative at this intersection by assumption. By the previous lemma, $\theta_{in}$ cannot change sign along the ingoing geodesic and therefore it stays negative until $p$. 
\end{proof}
\noindent\textit{Lemma 4. Under the assumptions of the theorem, an outgoing null geodesic initially inside a trapped region cannot leave it.}

\begin{proof}
By Lemma 2, the outgoing expansion cannot change sign along the outgoing null geodesic. The only possibility for the outgoing null geodesic to leave the trapped region is, therefore, that the ingoing null expansion changes sign. This is forbidden by Lemma 3~implying that the outgoing null geodesic cannot leave the trapped region.  
\end{proof}

We have now all the ingredients we need to prove the theorem.

\paragraph*{Proof of Theorem 2.} The proof of the theorem follows immediately from the lemmas just proved. Because of Lemma 4, the outgoing null geodesic cannot leave the trapped region. In turn, because of Lemma 1, the null convergence conditions are satisfied at all points in the future of $\mathcal{T}$ along the outgoing null geodesic. Therefore, all the assumption of the one sided Penrose theorem are satisfied and we can deduce that the spacetime is future null geodesically incomplete. \qed

\section{Discussion}
In this work we proved an incompleteness theorem for quasi-topological theories of gravity. We showed that under realistic initial conditions, the gravitational collapse in quasi-topological gravities must lead to a geodesically incomplete spacetime, provided matter satisfies the null energy condition and $h'(\psi)>0$ for $\psi>0$. The result applies to both polynomial theories in $D\geq5$ and their non-polynomial generalization in $D=4$ dimensions.

To the author's knowledge, the positivity condition $h'(\psi)>0$ for $\psi>0$ is satisfied by all quasi-topological theories known to produce non-singular black holes.  The nonnegative coefficients $c_n$ provide a simple sufficient condition for the positivity required by our theorem. This condition is satisfied by the explicit families constructed in Ref.~\cite{Bueno:2024dgm} and studied in Ref.~\cite{Bueno:2024zsx}. Within the convergence domain, these theories obey $h'(\psi)\geq1$ for $\psi\geq0$. The theorem also applies to theories with coefficients of either sign, provided their resummed characteristic function satisfies $h'(\psi)>0$ on the relevant positive domain.

Of course, quantum effects can violate the energy conditions and provide a way out of the theorem. However this is also true in general relativity~\cite{DiFilippo:2025kzh,DelPorro:2026njz}. 

In conclusion, quasi-topological theories admit vacuum black holes solution without curvature singularities. This might provide a useful framework for investigating their dynamics, including the generality of the mass inflation instability~\cite{Frolov:2026rcm,DiFilippo:2026jpv,Bueno:2026wjt}. 

Nevertheless, removing curvature divergences does not restore the predictivity of gravitational collapse and these theories are no better than general relativity at providing a complete and predictive description of black hole interiors.

\section*{Acknowledgments}
\noindent I acknowledge financial support from the European Union’s Horizon 2020 research and innovation programme under the Marie Sklodowska-Curie Actions (grant agreement ID:101207584)

\bibliographystyle{utphys}
\bibliography{references}
\end{document}